\documentstyle[prl,aps,epsf,multicol]{revtex}
\begin{document}
\draft
 
\title{Superconducting ``metals" and ``insulators"}
\author{S. Vishveshwara$^1$, T. Senthil$^2$, and Matthew P. A. Fisher$^2$}
\address{$^1$Department of Physics, University of California, Santa Barbara, CA 93106 \\
$^2$Institute for Theoretical Physics, University of California,
Santa Barbara, CA 93106--4030
}

\date{\today}
\maketitle

\begin{abstract}
We propose a characterization of zero temperature phases in disordered superconductors on the basis 
of the nature of quasiparticle transport. In three dimensional systems, there are two 
distinct phases in close analogy to the distinction between normal metals and insulators:
the superconducting ``metal" with delocalized quasiparticle excitations and the 
superconducting ``insulator" with localized quasiparticles. We describe 
experimental realizations of either phase, and study their general properties
theoretically. We suggest experiments where it should be possible to tune 
from one superconducting phase to the other, thereby probing a novel 
``metal-insulator" transition {\em inside} a superconductor. We point out various implications
of our results for the phase transitions where the superconductor is destroyed at 
zero temperature to form either a normal metal or a normal insulator.
\end{abstract}
\vspace{0.15cm}


\begin{multicols}{2}
\narrowtext

\section{Introduction}
While Cooper pairing in a superconductor is usually associated 
with a gap in the energy spectrum, it is quite well-known that the gap is 
by no means indispensable. Indeed, there are a number of situations in which 
superconductivity occurs with no gap in the quasiparticle excitation spectrum.
This possibility was first discussed in pioneering work by Abrikosov and Gorkov\cite{dG}
for $s$- wave superconductors in the presence of magnetic impurities. Another
novel instance is provided by Type $II$ $s$-wave superconductors in 
strong magnetic fields close to, but smaller than, $H_{c2}$.  
Yet another example of gapless superconductivity, of much current
interest, is the $d_{x^2 - y^2}$ superconductor; the d-wave characteristics of 
the pairing is reflected in four nodes in the gap function along which 
gapless quasiparticle excitations result. In all cases, 
the presence of low energy quasiparticle states has a profound 
effect on the low temperature thermodynamic and transport
properties of the superconductor. 

The effect of static disorder introduced by frozen impurities in such a 
gapless superconductor raises a number of fundamental questions. 
The low energy quasiparticles can either be delocalized and free to move 
through the sample, or can be localized by the disorder. These two
possibilities correspond to two
distinct superconducting phases that are distinguished by the nature of
quasiparticle transport. This distinction has a direct analogy
in the physics of normal (non-superconducting) systems where again
there are two possible phases distinguished by the nature of transport -
the metal with diffusive transport at the longest length scales, or the insulator
characterized by the absence of diffusion.  
Recent work\cite{short,dos}
has addressed the possible existence and properties of these superconducting phases 
in the context of dirty $d_{x^2 - y^2}$ 
superconductors. In particular, it was shown\cite{short} that quantum interference
effects destabilize the superconducting phase with delocalized quasiparticles in two dimensions.
Instead, the quasiparticle excitations in a two dimensional superconductor are
generically always localized\cite{note0}. 
In three dimensions however, both kinds of superconducting phases
are stable.

It is clear that the issues raised above are germane to
the properties of all superconducting systems.
Consider, for instance,
the case of dirty Type $II$ $s$-wave 
superconductors in strong magnetic fields. 
For fields above
$H_{c1}$, and in the absence of impurities, 
such a superconductor exists in the Abrikosov vortex lattice phase.
The quasiparticle excitation spectrum in the presence of a single 
isolated vortex (relevant when the field is just over $H_{c1}$) was shown by 
Caroli et. al.\cite{cmdg} to consist, at low energies,
of a set of discrete energy levels corresponding to states 
bound to the core of the vortex. The lowest lying excited state is separated 
from the ground state by an energy gap that is much smaller than the bulk gap,
but is nevertheless non-vanishing.
Impurities alter
this density of states, and could potentially 
close the (mini)gap. Furthermore,  the quasiparticle states
are no longer extended even along the direction of the vortex line due to
the strong effects of localization by the disorder in one dimension\cite{Zirn,short,dos}.
Impurities also
destroy the translational order of the Abrikosov vortex lattice, pinning the vortices
into a glassy phase.
When many
vortices are present, the quasiparticles can tunnel
from the core of one vortex to the other. The amplitude to tunnel 
depends on the overlap of the 
wavefunctions of the core states corresponding to
the individual vortices. This in turn is determined by the 
density and spatial configuration of the vortices. 
At low fields, the tunneling is insignificant, and the quasiparticle states 
are localized.

With increasing field however, the density of vortices,  
and hence the tunneling, increases. The nature of the quasiparticle states
- {\em i.e} whether they are delocalized or localized - is determined
by the interplay between tunneling and the disorder. One can envision 
two scenarios for what happens as the field is increased toward $H_{c2}$ at $T=0$.
One possibility is that the quasiparticle states remain localized all the way upto $H_{c2}$. We suggest that
this is the case if the non-superconducting ground state obtained for $H > H_{c2}$ is a localized insulator.
A second possibility is that the quasiparticle states undergo a delocalization transition
at some field $H_{c4} < H_{c2}$ which therefore precedes the destruction of superconductivity.
We suggest that this happens if the ground state at $H > H_{c2}$ is a normal metal. 

\begin{figure}
\epsfxsize=3.5in
\centerline{\epsffile{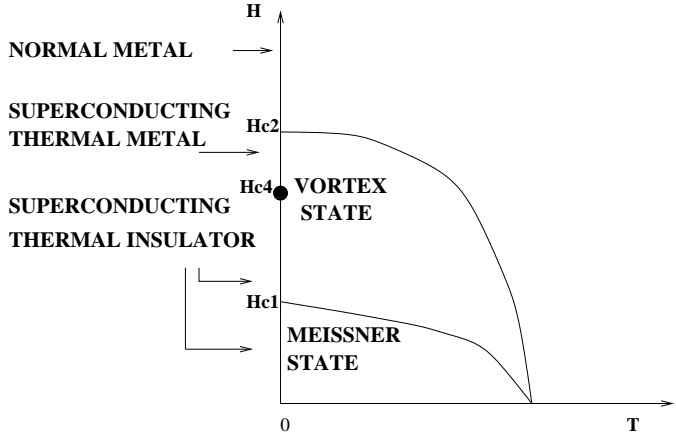}}
\vspace{0.15in}
\caption{A possible phase diagram for a dirty Type $II$ superconductor}
\vspace{0.15in}
\label{PD}
\end{figure}

Similar considerations apply as well to the situation where the superconductivity is 
destroyed at zero temperature and zero magnetic field, say, by increasing the amount of impurity scattering. 
If the transition is to a normal metal, it is likely to be from a superconducting  
phase with delocalized quasiparticle excitations. On the other hand, far from the transition 
in the superconducting phase,
the quasiparticle states at the Fermi energy, if any, will
be strongly localized. 
We expect therefore that a delocalization phase transition from this phase to the superconductor
with extended quasiparticle states  
precedes the ultimate transition from the superconductor to the normal metal. Again,
if the destruction of superconductivity leads to an insulating phase, it is natural 
to expect that the transition is directly from the superconductor with localized 
quasiparticle excitations.  

Thus the two kinds of superconducting phases, distinguished by the nature
of quasiparticle transport, are both realizable in a variety of
situations.
Despite the analogies with normal metals
and insulators,   
as pointed out in Ref.\cite{short} the problem of quasiparticle transport   
in a dirty superconductor is conceptually very different from the corresponding problem in a 
normal metal. This is because, unlike in a normal metal, the charge of the quasiparticles in a superconductor is {\em not} 
conserved, and hence cannot be transported through diffusion. However, 
the {\em energy} of the quasiparticles is conserved, and in principle the energy 
density could diffuse. In addition , in a singlet superconductor 
(in the absence of spin-orbit
scattering) the spin of the quasiparticles is a good quantum number and is conserved.  
Thus 
the quasiparticle 
spin density could also diffuse in the superconductor. 
We may therefore
characterize the two distinct kinds of superconductors by means of their 
energy (and spin) transport properties. 
The phase with delocalized quasiparticle excitations
has diffusive transport of energy (though not of charge) 
at the longest length scales. It is hence appropriate to call it a 
``superconducting thermal metal". In cases
where the quasiparticle spin is conserved, this phase also 
has spin diffusion at the longest length scales.
On the other hand, in the phase with localized 
quasiparticle excitations, there is no diffusion of energy or 
spin at the longest length scales.
Hence such a phase may be called a ``superconducting thermal insulator".

For the sake of
concreteness, we focus for most of the paper on the properties of 
dirty Type $II$ superconductors in strong 
magnetic fields. We discuss the possible phase diagrams, and analyse the properties of 
the two superconducting phases. This is first done in a BCS model of 
non-interacting quasiparticles, and assuming spin rotational invariance. 
In striking contrast to
normal (non-superconducting) systems,
quantum interference effects lead to singular corrections to the quasiparticle density 
of states in the superconductor\cite{dos}.  The density of states at the Fermi 
energy is non-zero in  
the superconducting ``metal'' phase, but has a  $\sqrt{E}$ cusp as a 
function of energy $E$ (measured from the Fermi energy). The 
superconducting ``insulator'' phase, on the other hand, has a vanishing density of states
at the Fermi energy. We will provide supporting numerical evidence for this 
result in agreement with earlier analytical calculations\cite{dos}.
Interactions between the quasiparticles could potentially modify the 
properties of either phase in important ways. These are analysed next.
In the ``metal'' phase, we show the existence of effects 
analogous to the Altshuler-Aronov singularities in a normal metal
due to the interplay of diffusion and interactions. Interaction
effects are more crucial in the ``insulator'' - we provide simple arguments 
showing that arbitrarily weak repulsive interactions lead to the formation of free 
paramagnetic moments. The ultimate fate of the ``insulator'' in the presence of
these free moments is a complicated problem, and we will not address it here. 
Weak attractive interactions however appear innocuous. 
We then conclude by discussing the many implications of this paper for 
experiments on various 
superconducting systems.

\section{Models}
In this section, we will introduce models appropriate 
to describe the physics of the quasiparticles
in a dirty superconductor in the pinned vortex phase.
We begin with a general BCS quasiparticle Hamiltonian:  
\begin{eqnarray}
{\cal H}_0 & = & \int_{x} c^{\dagger}_{\sigma}(x)\left(\frac{\left(-i\hbar\vec \nabla -\frac{e}{c}\vec A(x)\right)^2}{2m}
 - E_F + V(x) \right) c_{\sigma}(x)  \nonumber \\ 
& + & c^\dagger_{\uparrow} (x)\Delta(x)c^\dagger_{\downarrow} (x)  +  
c_{\downarrow} (x)\Delta^*(x) c_{\uparrow} (x) .
\label{HBCS}  
\end{eqnarray}
Here $m$ is the mass of the quasiparticles, and $E_F$ is the Fermi energy. 
The function $V(x)$ is a random potential due to impurities
that scatter the quasiparticles.  
The physical magnetic field is introduced through the vector potential $\vec A$. 
In the pinned vortex phase of a dirty superconductor, 
the gap function $\Delta(x)$ may be considered 
a random complex function of the position $x$ due to the 
random positions of the vortices. 
Both the vector potential and the complex gap function 
break time-reversal symmetry. As the phase
of $\Delta(x)$ winds by $2\pi$ on encircling a vortex, 
the corresponding term in the Hamiltonian that breaks 
time reversal symmetry  is of order the zero field gap $\Delta_0$.

	The spinful quasiparticles experience a shift in 
energy due to Zeeman splitting (not
included in the Hamiltonian Eqn.\ref{HBCS}). The largest field 
that we consider, $H_{c2}$, is of order 
$\frac{hc}{e}/\xi^{2}_{o}$, where $\xi_{o}$ is the 
coherence length, making the associated Zeeman energy $E_{zm}$ of 
order $\hbar^{2}/m\xi_{o}^{2}$. As $\xi_0$ is
related to $\Delta_0$ by $\Delta_0 \sim \frac{\hbar v_F}{\xi_0}$, we have
\begin{equation}
\frac{E_{zm}}{\Delta_0} \sim  \frac{\hbar}{p_F\xi_0} \sim 10^{-3} .
\end{equation}
Here $p_F = mv_F$ is the Fermi momentum. The last estimate uses the 
fact that the coherence length is typically a few thousand times larger
than the Fermi wavelength in conventional low-$T_c$ superconductors. Thus the 
Zeeman energy is negligible compared to the (zero field) gap, and 
we will ignore it for most of our discussion. We will also, for the most part,
assume the absence of any strong spin-orbit scattering. The system is then
spin rotational invariant. The Hamiltonian above also describes non-interacting
quasiparticles. Interaction effects are potentially quite important, and will
be discussed at some length later on in the paper.

It will often be convenient to think in terms of a 
lattice version of the Hamiltonian Eqn. \ref{HBCS}:
\begin{equation}
\label{HBCSL}
{\cal H}_{0L} = \sum_{ij}t_{ij}\sum_{\alpha}c^{\dagger}_{i\alpha}c_{j\alpha}
+ \left(\Delta_{ij}c^{\dagger}_{i\uparrow}c^{\dagger}_{j\downarrow} + h.c \right) .
\end{equation}
Hermiticity implies the condition $t_{ij} = t_{ji}^*$, and
spin rotation invariance requires $\Delta_{ij} = \Delta_{ji}$. 
Note that this is the most general spin rotational invariant Hamiltonian 
describing a superconductor with broken time reversal symmetry.
 
For some purposes, it is useful to use an alternate representation in terms of a new set of
$d$-operators defined by:
\begin{equation}
\label{c_to_d}
d_{i\uparrow} = c_{i\uparrow},~~ d_{i\downarrow} = c^{\dagger}_{i\downarrow } .
\end{equation}
The Hamiltonian, Eq. \ref{HBCSL}, then takes the form
\begin{equation}
\label{BCS_d}
{\cal H}_L = \sum_{ij} d^{\dagger}_i \left(\begin{array}{cc}t_{ij} & \Delta_{ij}
\\
                                                \Delta_{ij}^{*} & -t_{ij}^{*}
                                                               \end{array}
\right) d_j
                                     \equiv \sum_{ij}d_{i}^{\dagger}H_{ij}d_j .
\end{equation} 
Note that $SU(2)$ spin rotational invariance requires
\begin{equation}
\label{d_su2}
\sigma_y H_{ij}\sigma_y = -H^{*}_{ij} .
\end{equation}

The advantage of going to the $d$ representation is that the Hamiltonian
conserves the number of
$d$ particles. The transformation Eq. \ref{c_to_d} implies that the
number of
$d$ particles is essentially the $z$ component of the physical spin density:
\begin{equation}
\label{spin}
S^z_i = \frac{\hbar}{2}\left(d^{\dagger}_id_i -1 \right) .
\end{equation}
A spin rotation about the $z$ axis corresponds to a $U(1)$ rotation of the $d$
operators.
This $U(1)$ is clearly present in the $d$ Hamiltonian. Invariance under
spin rotations about the $x$ or $y$ axes is however not manifest.

The Hamiltonian may be diagonalized by the Bogoliubov transformation
\begin{eqnarray}
d_{i\uparrow}& = & \sum_n u_n(i)\gamma_{n\uparrow} + v_n^*(i)\gamma_{n\downarrow} , \\
d_{i\downarrow}& = & \sum_n v_n(i)\gamma_{n\uparrow} - u_n^*(i)\gamma_{n\downarrow} .
\end{eqnarray}
Here the $\gamma_{n\uparrow}, \gamma_{n\downarrow}$ are fermionic 
operators, and the functions $u_n(i), v_n(i)$ are determined
by solving the eigenvalue equation:
\begin{equation}
\label{BdG}
\sum_i H_{ij} \left[\begin{array}{c} u_n(j) \\
                                     v_n(j) \end{array}\right]
                         = E_n  \left[\begin{array}{c} u_n(i) \\
                                     v_n(i) \end{array}\right] .
\end{equation}
Note that these are just the familiar Bogoliubov-deGennes equations\cite{dG}.
The symmetry Eqn. \ref{d_su2} which follows from 
physical spin rotation invariance
then implies that
\begin{equation}
i\sigma_y\left[\begin{array}{c} u_n(i) \\
                                v_n(i) \end{array}\right] =  \left[\begin{array}{c} v_n^*(i) \\
                                    - u_n^*(i) \end{array}\right]
\end{equation}
is an eigenfunction of $H$ with eigenvalue $-E_n$. Thus the eigenvalues come in 
pairs $(E_n, -E_n)$. In terms of the $\gamma_n$ operators, the Hamiltonian
becomes
\begin{equation}
{\cal H}_L = \sum_n E_n \left(\gamma_n^{\dagger} \sigma_z \gamma_n \right) . 
\end{equation}
The ground state of the superconductor is obtained by filling all the negative
energy levels, {\em i.e} by occupying all the states created by
$\gamma^{\dagger}_{n\downarrow}$.

\subsection{Phase diagrams}
In this subsection, we discuss the zero temperature phase diagram of the dirty Type $II$ 
superconductor in the mixed state when the vortices are pinned by the disorder.
We are interested in 
characterizing the nature of quasiparticle transport
in such a superconductor. By analogy with normal (non-superconducting)
systems, we expect to distinguish two qualitatively different 
situations. The quasiparticle wavefunctions $(u_n, v_n)$ for
states at the Fermi energy (if any) may either be
extended through out the system, or they may be localized. 
Again, in analogy with normal metals, we expect that 
the phase with extended quasiparticle wavefunctions is
characterized by diffusive transport. However, as emphasized
in Ref. \cite{short}, the electric charge of the quasiparticles is not
conserved by the BCS Hamiltonian, and hence cannot diffuse.
The energy and spin densities of the quasiparticles are still conserved 
quantities and are thus capable of being transported through diffusion.
Thus the superconductor with extended quasiparticle states at the 
Fermi energy is characterized by diffusive transport of energy and spin
at the largest length scales. We will refer to this ground state as a
``superconducting thermal metal". Similarly, the phase with 
localized quasiparticle wavefunctions at the Fermi energy is 
characterized at $T=0$ by the absence of diffusion of energy and spin . 
We will refer to this phase as the ``superconducting thermal insulator". 

Recent studies of Anderson localization in a superconductor with spin rotation 
invariance
show that\cite{short} in two or lower dimensions, the quasiparticle wavefunctions are 
generically always localized. Above two dimensions however,  
phases with extended or localized states are possible. While these results are similar
to the corresponding results in a normal metal, there is a striking difference\cite{dos}: the quantum interference effects leading to localization give rise to singular corrections
in the quasiparticle density of states unlike in a normal metal. We will provide additional 
support for this result, and explore its 
consequences in later sections.

First consider the situation at low fields $H$ just bigger than $H_{c1}$
when the inter-vortex spacing is large.    
For a single isolated vortex, in the absence of impurities,
there exist states bound to the core\cite{cmdg}  
whose energies form a discrete set of minibands. (The (mini)bandwidth
is entirely due to motion along the vortex line).
There still is a gap to all excitations of order  $\frac{\Delta_0^{2}}{E_{f}}$. 
This is very small compared to $\Delta_0$, 
which in turn is of order $k_B T_{c}$. In the presence of impurities, the quasiparticle wavefunctions 
are localized along the direction of the vortex line\cite{Zirn,short,dos}. 
Further, for large inter-vortex spacing ,
quasiparticle tunneling from the core of one vortex 
to another is insignificant. Thus we expect
no energy or spin transport by the quasiparticles at the longest length scales; 
the system is in the thermal insulator phase.

\begin{figure}
\epsfxsize=3.5in
\centerline{\epsffile{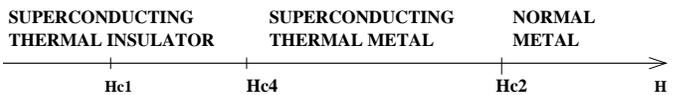}}
\vspace{0.15in}
\caption{A possible phase diagram for a dirty Type $II$ superconductor}
\vspace{0.15in}
\label{SCNM}
\end{figure}

Upon increasing the magnetic field towards $H_{c2}$,  it becomes
important to include tunneling from the core of one vortex to the other.
If the tunneling is sufficiently strong, it may become possible for 
the quasiparticle states to delocalize, leading to a thermal metal phase. 
Consider, in particular, the situation where the ultimate destruction of 
the superconductivity at $H_{c2}$ leads to a normal metal. 
The normal metal is characterized by diffusive transport of charge, spin,
and energy at the longest length scales. It is natural to expect that generically, 
the spin and thermal diffusion already exist in 
the superconducting phase just below $H_{c2}$. 
In other words,  
we expect that a 
transition to a normal metal at $H_{c2}$ occurs from the superconducting thermal metal phase. 
Thus, 
in this case, there are three distinct zero temperature phases as the 
magnetic field is increased from just above $H_{c1}$
to just above $H_{c2}$. The superconducting thermal
insulator at low fields (above $H_{c1}$) first undergoes a 
delocalization 
transition to the superconducting thermal metal  at some field $H_{c4} < H_{c2}$ 
before the superconductor is destroyed to form the normal 
metal (see Figure \ref{SCNM}).

\begin{figure}
\epsfxsize=3.5in
\centerline{\epsffile{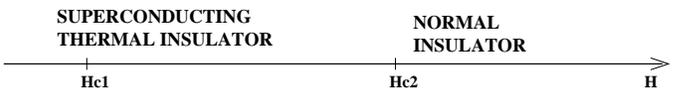}}
\vspace{0.15in}
\caption{Another possible phase diagram for a dirty Type $II$ superconductor}
\vspace{0.15in}
\label{SCI}
\end{figure}

In the other case where the destruction of superconductivity 
at $H_{c2}$ leads to a localized insulator, it is natural to expect that
the transition occurs from a superconductor where the quasiparticles are localized, {\it i.e} 
directly from the superconducting thermal insulator  (see Figure \ref{SCI}). 

\section{Superconducting thermal metal}
\label{SSM}
In this section, we examine the properties of the superconducting thermal metal phase in
more detail. In this phase, there is diffusive transport of spin and energy. 
The quasiparticle density of states at the Fermi energy is non-zero and finite.

We may quantify the spin transport by defining a ``spin conductivity" in analogy with the electrical conductivity for charge transport. The role of the chemical potential is
played by a Zeeman magnetic field $B$ coupling to, say, the $z$-component of the spin. The analog 
of the electric field is thus the spatial derivative of the Zeeman field. 
The spin conductivity $\sigma_s$
thus measures the $z$ component of the 
spin current $\vec j^z_s$ induced in the system in response to an externally applied
spatially varying Zeeman field along the $z$-direction of spin:
\begin{equation}
\vec j^z_s = -\sigma_s g\mu_B\vec \nabla B .
\end{equation}
Here $g$ is the gyromagnetic ratio, and $\mu_B$ the Bohr magneton.
It is easy to show that $\sigma_s$ satisfies an Einstein relation
\begin{equation}
\label{ES}
\sigma_s =  \frac{D_s \chi_0}{(g\mu_B)^2} ,
\end{equation}
where $D_s$ is the spin diffusion constant, and $\chi_0$ is the spin susceptibility.
In the approximation of ignoring quasiparticle interactions,
$\chi_0$ is simply proportional to the
quasiparticle density of states $\rho_0$ at the 
Fermi energy:
\begin{equation}
\chi_0 = \frac{(g\mu_B \hbar)^2}{4}\rho_0 .
\end{equation}
 In the thermal metal phase, 
$\sigma_s$ is finite and non-zero at zero temperature.

In the vortex phase, the 
diffusion parallel to the magnetic field, described by the diffusion constant $D_{\parallel}$, is mainly along the core 
of the vortices, while the diffusion perpendicular to it, with diffusion constant $D_{\perp}$, is due to in-plane motion 
between vortices. The latter depends strongly on the inter-vortex 
tunneling strength, and in general we expect the diffusion to be highly anisotropic.  
For ease of presentation, we will, for the time-being, 
assume that the diffusion is isotropic. When appropriate, we will take into account 
the anisotropic diffusion.

The energy diffusion is measured by the more familiar thermal conductivity $\kappa$. 
This too satisfies an
Einstein relation:
\begin{equation}
\label{ET}
\kappa = D_T C ,
\end{equation}
where $D_T$ is the thermal diffusion constant, and $C$ the specific heat. In the approximation
of ignoring quasiparticle interactions, the specific heat is determined by the 
density of states. In particular, in the limit $T \rightarrow 0$, where $T$ denotes the temperature, we have
\begin{equation}
C = \frac{\pi^2}{3}k_B^2 \rho_0 T .
\end{equation}
Furthermore, in the non-interacting theory, as both spin and energy transport
are by the quasiparticles, the corresponding diffusion constants are the same:
\begin{equation}
D_s = D_T .
\end{equation}
This then leads to a ``Weidemann-Franz" law relating the spin and thermal conductivities:
\begin{equation}
\frac{\kappa}{T\sigma_s} = \frac{4\pi^2 k_B^2}{3\hbar^2} .
\end{equation}

\begin{figure}
\epsfxsize=3.5in
\centerline{\epsffile{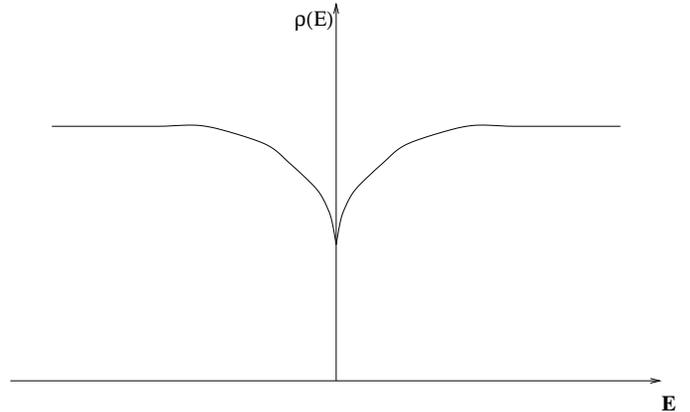}}
\vspace{0.15in}
\caption{Quasiparticle density of states of the superconducting thermal metal }
\vspace{0.15in}
\label{metdos}
\end{figure}

Though the quasiparticle density of states is finite and non-zero at the 
Fermi energy, as we show below, it varies as $\sqrt{E}$ on moving in energy ($E$) away
from the Fermi energy (See Fig. \ref{metdos}). This is in sharp contrast to a normal metal of non-interacting
electrons , and has its origins in quantum interference effects specific to the 
superconductor. This result can be established in a field theoretic analysis of 
quantum interference effects in the superconducting thermal metal  phase\cite{dos}. Here, we provide instead a simple physical explanation of the effect using a semiclassical 
argument developed earlier 
in a different context\cite{BB}.

The symmetry Eqn. \ref{d_su2} of the Bogoliubov-deGennes Hamiltonian implies that the amplitude 
$iG_{ij,\alpha \beta} = \langle i\alpha |e^{-\frac{iHt}{\hbar}}|j\beta\rangle \theta(t)$ for a 
$d$-particle to go from point $j$, (pseudo)spin $\beta$ , to point $i$, spin $\alpha$, 
satisfies the relations:
\begin{eqnarray}
\label{Gsymm_1}
G_{ij,\uparrow \uparrow}(t) &= -G^*_{ij,\downarrow \downarrow}(t) ,\\
\label{Gsymm_2}
G_{ij,\uparrow \downarrow}(t) &= G^*_{ij,\downarrow \uparrow}(t) .
\end{eqnarray}
The Fourier transform of this amplitude is given by
\begin{eqnarray*}
G_{ij,\alpha\beta}(\omega +i\eta) & = & \int dt e^{i(\omega + i\eta)t} G_{ij, \alpha\beta}(t) \\
& = &  \langle i\alpha |\frac{1}{\frac{\hbar \omega - H}{\hbar} + i\eta}|j\beta\rangle  .                                
\end{eqnarray*}
The quasiparticle density of states at an energy $E$ away from the Fermi energy 
may be obtained from this in the usual manner:
\begin{equation}
\rho(E) = -\frac{1}{\pi \hbar}Im\left(\overline{G}_{ii,\uparrow\uparrow}(E + i\eta) + (\uparrow 
\leftrightarrow \downarrow) \right) ,
\end{equation}
where the overbar denotes an ensemble average over impurity configurations.Consider now the return amplitude $G_{ii,\uparrow \uparrow}(t)$.
This can be written as a sum over all possible paths for this event. 
Consider in particular the contribution from the special class of paths
where the particle traverses some orbit and returns to the point $i$ in time $t/2$
with spin down, and then traverses the same orbit again in the remaining time and returns 
with spin up. Using the symmetry relation Eqn.\ref{Gsymm_2}, this contribution to $iG_{ii,\uparrow \uparrow}(t)$ can be written as
\begin{displaymath}
iG_{ii,\uparrow \downarrow}\left(\frac{t}{2}\right)iG_{ii,\downarrow \uparrow}\left(\frac{t}{2}\right)~~
= ~~-|G_{ii,\uparrow \downarrow}\left(\frac{t}{2}\right)|^2 .
\end{displaymath}
Now 
$|G_{ii,\uparrow \downarrow}(\frac{t}{2})|^2$ is just the probability 
for the event $i\uparrow \rightarrow i\downarrow$ in time $t/2$. For large $t$, this is 
half the total return probability $P(t)$. This in turn is determined by the condition that the 
$d$-particles diffuse through the system. For diffusing particles in three dimensions,
the return probability is
\begin{equation} 
P(t) =  \frac{1}{ (4\pi D_s t)^{3/2}} .
\label{diffn}
\end{equation}
This then leads to an energy dependent correction to the density of states:
\begin{equation}
\label{rhossm}
\rho(E)- \rho_0  = \frac{1}{  2\pi^2(D_s \hbar)^{3/2}}\sqrt{E} .
\end{equation}

If we take into account the anisotropic nature of the diffusion in 
the vortex phase, then the same result holds, but with an effective 
diffusion constant $D_s = \left( D_{\perp}^2 D_{\parallel} \right)^{\frac{1}{3}}$.

The energy dependence of the density of states has important consequences for the 
low temperature thermodynamics of the superconducting thermal metal phase. In particular, the
specific heat at low temperature behaves as
\begin{equation}
C = \gamma T + b_{cq}\left(\frac{k_BT}{2\pi\hbar D_s}\right)^{3/2} ,
\end{equation}
where $\gamma = \frac{\pi^2}{3}k_B^2 \rho_0$, and the constant factor is given by
\begin{equation}
b_{cq} = \frac{15 k_B}{(2 )^{3/2}}\left(1 - \frac{1}{2^{3/2}}\right)\zeta(5/2) .
\end{equation}
Here $\zeta(5/2) = \sum_n \frac{1}{n^{5/2}}$. Note that $b_{cq}$ is a universal constant.    
Similarly, the spin susceptibility at low temperatures behaves as
\begin{equation}
\chi(T) - \chi_0 =  \frac{b_{sq}}{D_s} \left(\frac{k_B T}{2\pi \hbar D_s}\right)^{1/2} .
\end{equation}
The constant $b_{sq}$ is again universal: 
\begin{equation}
b_{sq} = \frac{(g\mu_B)^2 \hbar {\cal I}}{(2\pi)^{3/2}} .
\end{equation}
with ${\cal I} = \int_{0}^{\infty} \frac{dx}{\sqrt{x}} \left(\frac{1}{e^x + 1}\right) \approx 1.07$.

It is amusing to note that this correction has the same form as the Altshuler-Aronov effects in a diffusive,
interacting normal metal, though the physics is quite different. Later on in the paper, when we consider
interaction effects , we will show the existence of an Altshuler-Aronov correction of the 
same form in the superconducting thermal metal as well.

\section{Superconducting Thermal Insulator}
\label{SSI}
We now consider the properties of the superconducting thermal insulator phase. By definition, this is 
a superconductor where  the thermal
conductivity has the limiting form $\kappa/T \rightarrow 0$ as $T \rightarrow 0$. Similarly,
the spin conductivity is given by $\sigma_s = 0$ at zero temperature. Thus this phase is a superconductor
for charge transport, but an insulator for thermal and spin transport. 

In contrast to conventional disordered
insulators, the density of quasiparticle states actually {\em vanishes} in the superconducting 
thermal insulator. This can be seen by the following simple argument\cite{dos}.  
Consider the Hamiltonian (\ref{HBCSL}) in the limit of strong on-site randomness and weak
hopping between sites. In the extreme limit of zero hopping, the sites are all decoupled. 
At each site, the Hamiltonian in terms of the $d$-particles satisfies
the $SU(2)$ invariance requirement $\sigma_y H \sigma_y = -H^*$. This constrains the Hamiltonian
to be of the form
$H =  t\sigma_z + \Delta_r \sigma_x + \Delta_i \sigma_y$ with $t, \Delta_r, \Delta_i$ random. 
Physically, $t$ can be thought of as a random on-site chemical potential, and 
$\Delta_r, \Delta_i$ as the real and imaginary parts of the random on-site BCS order parameter $\Delta$.
Considering now the case where the joint probability distribution
of $t, \Delta_r, \Delta_i$ has finite, non-zero weight at zero, we see immediately that the disorder
averaged density of states vanishes as $E^2$.  Now consider weak non-zero hopping. In the localized phase,
perturbation theory in the hopping strength should converge, and we expect to recover the 
single site results at asymptotically low energies. If the joint probability distribution of $t, \Delta_r, \Delta_i$
has vanishing weight at zero (as happens for instance, for $\Delta_r, \Delta_i$ 
non-random and non-zero, and only $t$
random), then the density of states only vanishes even faster. To get a density of states that vanishes 
slower than $E^2$, or is finite at the Fermi energy requires a diverging probability density at 
$t = \Delta_r = \Delta_i =0$ which is presumably unphysical, and definitely non-generic. Thus, we
conclude that the (disorder averaged) quasiparticle density of states vanishes in the superconducting thermal
insulator phase in the absence of quasiparticle interactions. 

We now demonstrate the validity of this argument by 
direct numerical calculation of the density of 
eigenstates
of the Bogoliubov-deGennes equations in the spin insulating phase. 
The simulations were done in one spatial dimension.
The advantage of doing so is three-fold: as localization effects are strongest in one
dimension, it is easier to access the properties of the localized phase in $d = 1$. Further,
it is possible to go to fairly large system sizes in $d = 1$ and hence the results are more reliable.
Finally, we expect that the properties of the localized 
phase are qualitatively the same in any dimension.
Hence it is sufficient to consider the $d = 1$ case. 
A physical realization of a one dimensional system
to which our results are directly applicable is obtained by 
considering the quasiparticle states in the 
core of a single, isolated vortex in the presence of disorder.

To simulate the density of states of the d-particles, we employ the Hamiltonian in Eqn. \ref{BCS_d}:
\begin{equation}
{\cal H}=\sum_{i,j}d_{i}^{\dagger}\left(t_{ij}\sigma_{z} + \Delta_{ij}^{(1)}\sigma_{x} + \Delta_{ij}^{(2)}\sigma_{y}\right)d_{j} ,
\label{ham}
\end{equation}
where now the sites $i$ and $j$ reside on a one-dimensional lattice with periodic boundary conditions.  
It is convenient to picture the Hamiltonian in terms of coupled spin-up and spin-down sublattices.

We now employ the Hamiltonian in Eq.\ref{ham} to explore the density 
of states numerically for various models and degrees 
of disorder. We  begin with a model which shows a gap in the density of states in the absence of disorder.  We  
set the nearest neighbour coupling $t_{i i+1}$ to a constant, $t$, and take  $\Delta = \delta_{ij}\Delta^{\small 0}$ to be on-site and real. The pure Hamiltonian can be diagonalized trivially. The resulting
single particle excitations have a dispersion 
\begin{equation}
E = \pm\sqrt{(\Delta^{{\small 0}})^{2} + 4 t^{2}\cos{k}} ,
\label{purgap}
\end{equation}
with a gap $2 \Delta^{\small 0}$ about $E = 0$, and a bandwidth $2\sqrt{(\Delta^{{\small 0}})^{2} + 4 t^{2}}$ . We now introduce disorder by allowing the on-site couplings $t_{ii}$, 
$\Delta_{i i}^{(1)}$ and $\Delta_{i i}^{(2)}$ to take the values $V^1$, $\Delta^0 + V^2$ and $V^3$ respectively. 
The $V^i$ are random variables drawn from a uniform distribution with zero mean and width $W$ which acts as
a measure of the disorder strength.  Note that a non-zero value of $\Delta^{(2)}$  breaks time-reversal invariance, 
as one would expect for the vortex phase.

\begin{figure}
\epsfxsize=3.5in
\centerline{\epsffile{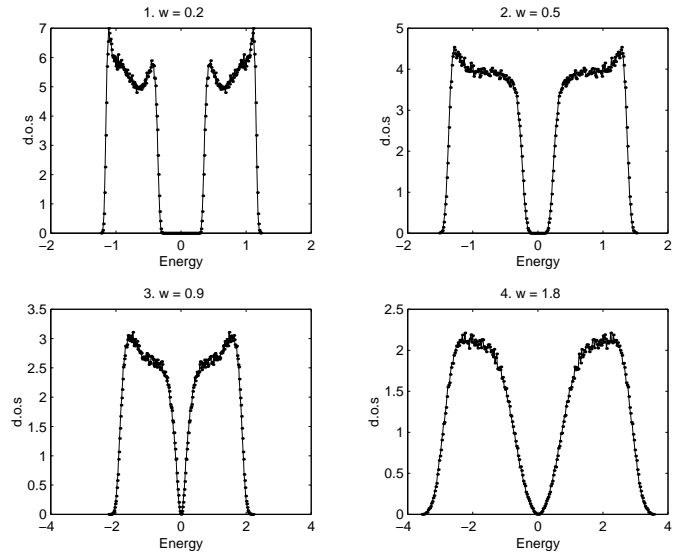}}
\vspace{0.15in}
\caption{Density of states in the superconducting thermal insulator showing 
evolution with increased disorder, W. The energy scale is set by the 
bandwidth of the pure dispersion in Eq.\ref{purgap}.}
\vspace{0.15in}
\label{noTgap}
\end{figure}

\begin{figure}
\epsfxsize=3.5in
\centerline{\epsffile{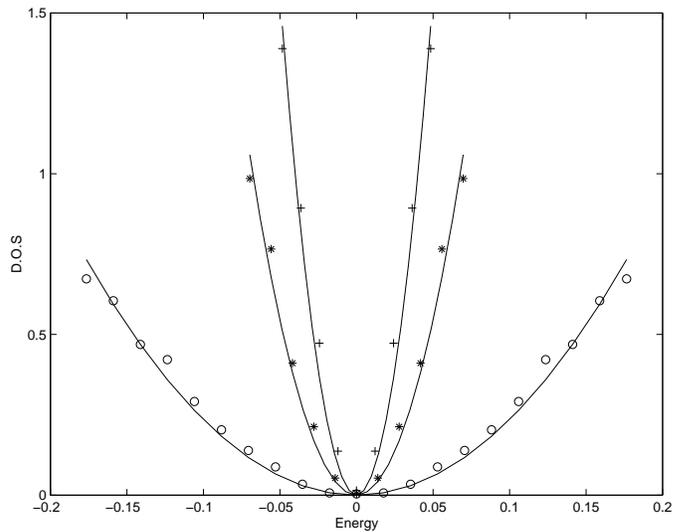}}
\vspace{0.15in}
\caption{Density of states of the superconducting thermal insulator at low energies. 
Here, the constant 'A' of Eq.\ref{noTdos} decreases with increasing disorder strength W that ranges from 0.8 to 2. }
\vspace{0.15in}
\label{zoomnoT}
\end{figure}

As seen in Fig.\ref{noTgap}, impurities begin to fill in the gap. (The precise manner 
is specific to the distribution of disorder, which 
we verified by using different forms for the probability distribution of the $V^i$). 
One can observe the symmetry about $E = 
0$ which is a result of the particle-hole symmetry of the Bogoliubov deGennes Hamiltonian.  
As the disorder strength is increased, there is an increasing density of states in the gap. But the density of states
at the Fermi energy nevertheless always vanishes. Closer examination of the density of states near the Fermi energy
(see Figure \ref{zoomnoT})
shows that it actually vanishes as
\begin{equation}
\rho(E) = A |E|^2,
\label{noTdos}
\end{equation}
where A is a constant. This power law is exactly what is predicted by the simple argument outlined at the beginning of
this section.

\begin{figure}
\epsfxsize=3.5in
\centerline{\epsffile{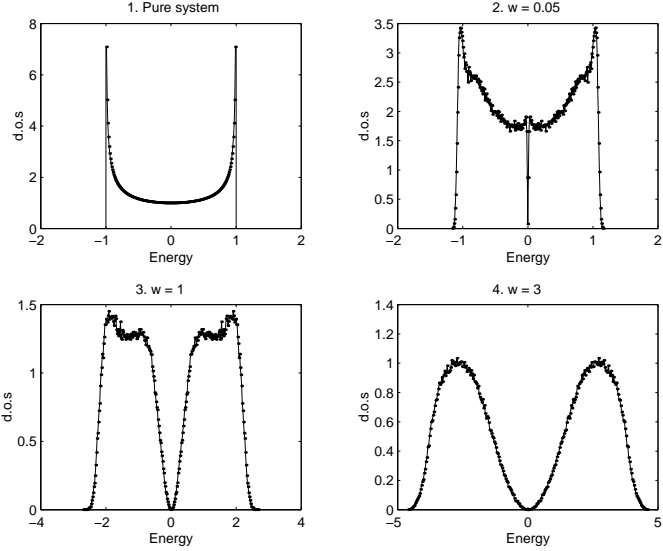}}
\vspace{0.15in}
\caption{Evolution of the density of states with disorder ,W, for a model where the density of states is constant in the clean limit. Here the energy scale is set by the bandwidth specified by the pure dispersion of Eq.\ref{purconst}.} 
\vspace{0.15in}
\label{noTconst}
\end{figure}

We now consider the situation where the density of states is a constant in the absence of 
disorder. We set the nearest neighbor coupling $t_{ii+1}$ and gap $\Delta_{ii+1}$ to be 
real constants t and $\Delta$ respectively. We then obtain for the dispersion of the single particle excitations the form
 \begin{equation}
E = 2\sqrt{\Delta^{2} +  t^{2}}\cos{k} ,
\label{purconst}
\end{equation}
with a bandwidth $4\sqrt{\Delta^{2} +  t^{2}}$. 
We introduce disorder by letting the on-site couplings 
$t_{ii}$ , $\Delta^{(1)}_{ii}$ and $\Delta^{(2)}_{ii}$ take the 
values $V^1$, $V^2$ and $V^3$ respectively, where the $V^i$'s are 
random variables as specified for the case with the pure gapped dispersion.
As seen in Fig. \ref{noTconst}, disorder reduces the density of states at 
the Fermi energy, ultimately forcing it to vanish as $E^2$.

In summary, the superconducting thermal insulator phase has 
the remarkable feature that the quasiparticle
density of states actually vanishes at the Fermi energy. This is in striking contrast
to a conventional Anderson insulator, and has several obvious consequences for the 
low temperature thermodynamic properties of the phase. 

The vanishing density 
of states also has consequences for thermal and spin transport at finite temperature which is
presumably through variable-range hopping. This can be seen as follows:
for a hop between two localized quasiparticle states separated by a distance $R$,
the overlap of the two states $\sim e^{-\frac{R}{\xi}}$ where $\xi$ is the localization length.
The typical energy separation $E_R$ between these states is 
determined by the density of states. When this vanishes as $E^2$, 
the total number of states in a radius $R$ in an energy interval $E$ about the Fermi energy
is of order $E^3 R^3$ (in three dimensions). 
Therefore the typical spacing $E_R \sim \frac{1}{R}$.
The total rate for hopping a distance $R$ is proportional to $e^{- (\frac{2R}{\xi} + \frac{c}{RT})}$.
The first term in the exponential is the square of the wave-function overlap, and the second 
is the activation energy cost for the hop. (Here $c$ is some unknown constant determined by the density of states).
The total hopping rate, as measured 
by the spin or thermal conductivities, is obtained by summing over hops of all possible 
distances $R$. For low temperature, it is clear that the sum will be dominated by hops of 
size $R \sim \frac{1}{\sqrt{T}}$. Thus the total hopping rate (and hence the spin/thermal 
conductivities) $\sim e^{-\sqrt{\left(\frac{T_0}{T}\right)}}$. Note that the exponent $1/2$ 
differs from the conventional Mott exponent of $1/4$, and is due to the density of states
vanishing as $E^2$.

\section{Interactions}
\label{Int}
Our discussion of the properties of the two superconducting phases
has thus far been based on the non-interacting quasiparticle Hamiltonian of
Eqn. \ref{HBCS} or Eqn. \ref{HBCSL}. In this section, we consider the effects 
of interactions between the quasiparticles. A discussion of the 
microscopic origin of these interactions is provided in Appendix \ref{intham}. Here we
take a more phenomenological approach. It is important to 
note that the interactions are short-ranged in space - the long-range
Coulomb repulsion between the underlying electrons is screened out by the 
condensate. We keep only the interactions in the triplet channel
to get the Hamiltonian
\begin{eqnarray}
{\cal H} & = & {\cal H}_0 + {\cal H}_{int} ,\\
{\cal H}_{int} & = & \int d^d x   u_t(x-y)\vec S(x) \cdot \vec S(y) ,
\label{Hint}
\end{eqnarray}
where $ u_t$ is short-ranged, and $ \vec S(x)$ is the spin
density. Note that as the charge density is not a hydrodynamic mode, 
the singlet interaction
is expected to be quite innocuous, and we simply drop it. As
discussed in Appendix \ref{intham}, 
the sign of $u_t$ could be either positive or negative, with
negative $u_t$ corresponding to repulsive interactions.   

Interactions can also be included instead in the lattice model Eqn. \ref{HBCSL}. We expect that 
the universal properties of the two superconducting phases are insensitive to the detailed form 
of the interaction Hamiltonian. So we may consider any short-ranged interaction with the right 
symmetries (spin conservation). A particularly simple choice is provided by an on-site
Hubbard interaction:
\begin{eqnarray}
{\cal H}_L & = & {\cal H}_{L0} + {\cal H}_{U} , \\
{\cal H}_{U} & = & \frac{U}{2}\sum_i n_i(n_i -1) ,
\label{Hubb}
\end{eqnarray}
where $n_i = c_i^{\dagger}c_i$ is the number operator for site $`i'$. For repulsive interactions, $U > 0$.  

In the rest of this section, we will use these model interaction Hamiltonians to 
discuss their effects on the two superconducting phases.

\subsection{Superconducting thermal metal}
Interaction effects lead to significant changes in the properties of 
diffusive normal metals\cite{AA}, as was shown by Altshuler and Aronov.
In the superconducting thermal metal phase, it is natural to expect similar effects 
due to the interplay of spin diffusion and interactions. We demonstrate this below
with some simple calculations.  

Consider the free energy of the interacting quasiparticles in the superconducting thermal metal. 
To calculate this, it is convenient to pass to a functional integral
representation for the partition function:
\begin{eqnarray}
Z & = & \int \left[{\cal D}\psi \right] e^{-\left(S_0 + S_{int}\right)} ,\\
S_0 & = & \int d \tau d^d x ~~(\bar{\psi}(x,\tau)\frac{\partial}{\partial \tau}\psi(x,\tau) 
+ {\cal H}_0(\bar{\psi},\psi)) , \\
S_{int} & = & \int d\tau d^dx d^dy~~ u_t(x -y) \vec S(x,\tau)\cdot \vec S(y, \tau)
\end{eqnarray}
As we show below, all the singular corrections to
the free energy come from the diffusive nature of the spin 
fluctuations. (The charge density does not diffuse, and hence the singlet interaction does not
contribute to any singular corrections ;we are justified in dropping it). 
The first order correction to the free energy due to interactions is given by
\begin{eqnarray}
V\beta \Delta{\cal F} & = & \langle S_{int} \rangle \nonumber \\
& = &  \int d\tau d^dx d^dy~~ u_t(x -y)\overline{ \langle \vec S(x,\tau)\cdot \vec S(y, \tau) \rangle} ,
\end{eqnarray}
where $V$ is the system volume, $\beta$ the inverse temperature, 
${\cal F}$ the free energy density, and $\Delta {\cal F}$ the correction
to the free energy density due to interactions. The expectation value 
on the right hand side is to be evaluated in the non-interacting theory. The overline denotes
an average over all realizations of the disorder.
The expectation value can be related directly to the spectral weight for 
spin density fluctuations as follows:
\begin{eqnarray*}
\langle \vec S(x,\tau)\cdot \vec S(y, \tau) \rangle & = & \langle  \vec S(x,0)\cdot \vec S(y, 0) \rangle \\
& = & \sum_{nm} \frac{e^{-\beta E_n}}{Z_0} \langle n|\vec S(x)|m\rangle \cdot \langle m|\vec S(y)|n \rangle ,
\end{eqnarray*}
where $|n\rangle, |m\rangle$ are exact eigenstates (with energies $E_n, E_m$ respectively)
of the non-interacting Hamiltonian ${\cal H}_0$, and
$Z_0 = Tr e^{-\beta {\cal H}_0}$ is the corresponding partition function. Now the spectral
weight for spin density fluctuations in the non-interacting theory can be expressed as
\begin{eqnarray*}
\chi''(x,y;\omega) & = & \sum_{nm}\frac{e^{-\beta E_n}}{Z_0} 
\langle n|\vec S(x)|m\rangle \cdot \langle m|\vec S(y)|n \rangle \\
&   & (2\pi) \delta(\omega - \frac{(E_m - E_n)}{\hbar})\left(1 - e^{-\beta\hbar\omega}\right) .
\end{eqnarray*}
Clearly then, we have
\begin{equation}
\langle \vec S(x,0)\cdot \vec S(y, 0) \rangle = \int_{-\infty}^{\infty} \frac{d\omega}{2\pi}
\frac{\chi^{''}(x,y; \omega)}{1 - e^{-\beta\hbar\omega}} .
\end{equation}
Upon averaging over the disorder, translation invariance is restored, and we have
\begin{equation}
\overline{\chi^{''}(x,y; \omega)} = \int \frac{d^dq}{(2\pi)^d} e^{i\vec q. (x-y)}\chi^{''}(q, \omega) .
\end{equation}
The singular contributions to the free energy all come from the small $q, \omega$
behaviour of $\chi^{''}$. This is entirely determined by the diffusive nature of spin transport,
\begin{equation}
\chi^{''}(q, \omega) = \frac{6\hbar \sigma_s \omega q^2}{\omega^2 + D_s^2q^4},
\label{specwt}
\end{equation}
where $\sigma_s$ is the spin conductivity  and $D_s$ is the spin diffusion 
constant. (There is an extra factor of $3$ in the expression above from the
standard form ,which is due to summing over all three
components of the spin). We can now evaluate the free energy correction 
(See Appendix  \ref{AAC} for details). The result is given by
\begin{equation}
\Delta{\cal F}(T) - \Delta{\cal F}(0) \sim -\tilde{u_t} T^{5/2}
\end{equation}
in three dimensions. (Here we have $\tilde{u_t} = \int d^d x u_t(x)$). 
This then leads to a correction to the specific heat due to interactions
of the form
\begin{equation}
\Delta C = b_{ci}\left(\frac{k_BT}{2\pi\hbar D_s}\right)^{3/2} , 
\end{equation}
where the coefficient $b_{ci}$ is given by
\begin{equation}
b_{ci} = -\frac{45}{8}\tilde{u_t}\frac{\chi_o}{(g\mu_B)^2} k_B \zeta(5/2) .
\end{equation}
Note that this is similar in form to the correction due to quantum interference
pointed out in Section \ref{SSM}, but has a different physical origin. 
(As before, with anisotropic diffusion, $D_s$ is replaced by the effective diffusion
constant $ \left(D_{\perp}^2 D_{\parallel}\right)^{\frac{1}{3}}$). The total
specific heat in the superconducting thermal metal phase therefore is
\begin{equation}
C = \gamma T + \tilde{b}_c \left(\frac{k_BT}{2\pi\hbar D_s}\right)^{3/2} ,
\end{equation}
where the constant $b_c$  has contributions from both quantum interference
and interaction effects:
\begin{equation}
b_c = b_{cq} + b_{ci} .
\end{equation}
Note that $b_{cq}$ is a universal number, while $b_{ci}$ depends on the interaction strength
and the zero temperature spin susceptibility. This should allow experimental determination of the 
importance of interaction effects by simultaneous measurements of the specific heat
and the spin or thermal diffusion constant. The deviation of the coefficient $b_c$ from the universal 
value $b_{cq}$ gives a measure of the interaction strength.

Similar effects exist in the spin susceptibility as well. To see this, consider evaluating
the free energy in the presence of an externally applied Zeeman magnetic field 
coupling to the $z$ component of the spin. This does not affect the diffusion of
$S_z$, but the diffusion of $S_x, S_y$ gets cut off. The precise manner in
which this happens can be found from hydrodynamic considerations by including the effect
of the precession of the spin density under the external magnetic field in the
classical diffusion equation. To that end, first note that in the presence of 
a small magnetic field  $B$ coupling to the $z$-component 
of the spin, the ground state has a spin density
\begin{equation}
\vec S = \hat{z} \frac{\chi_0}{g\mu_B} B .
\end{equation}
Consider small deviations of the local spin density from the ground state:
\begin{equation}
\vec S(x,t) = \left(\delta S_x(x,t), \delta S_y(x,t), \frac{\chi_0}{g\mu_B} B + \delta S_z(x,t)\right) .
\end{equation}
Quite generally, this satisfies the equation of motion
\begin{equation}
\frac{\partial \vec S}{\partial t} + \vec S \times (g\mu_B \vec B) = -\partial_i \vec j_s^i ,
\end{equation}
where $i = 1,2,3$ is a spatial coordinate index and $\vec j_s^i$ is the spin current vector
in the spatial direction $i$. The second term on the left hand side arises
from the precession of the spin under the external magnetic field. In the absence of this 
term, the equation above reduces to the familiar continuity equation expressing 
spin conservation. To derive the $S^+S^-$ correlator, introduce a small additional  
magnetic field $b(x,t)$ coupling to $S_x$. Then, the spin currents $j_s^{xi}, j_s^{yi}$
are related to gradients of the spin density and the field $b$ through 
\begin{eqnarray}
j_s^{xi} & = & -D_s \partial_i \delta S^x + (g\mu_B\sigma_s)\partial_i b  ,\\
j_s^{yi} & = & -D_s \partial_i \delta S^y ,
\end{eqnarray}
where $D_s, \sigma_s$ are the spin diffusion constant and spin conductivity respectively.
We may now determine the response of the spin density to the field $b$ from these equations:
\begin{equation}
\delta s^{+}(k, \omega)  = \frac{g\mu_B \sigma_s}{D_s}\left[\frac{D_sk^2 - i\tilde{B}}
{-i(\omega +\tilde{B}) + D_s k^2}\right] b(k, \omega) ,
\end{equation}
where $\tilde{B} = g\mu_B B$. We have used the Einstein relation Eqn.\ref{ES} to express $\chi_0$ in 
terms of $\sigma_s$ and $D_s$. A similar expression, but with $B \rightarrow -B$, holds for 
$\delta s^-(k, \omega)$. The response function $\chi_{xx}(k,\omega)$ can now be read off
to be
\begin{equation}
\chi_{xx}(k, \omega) = \frac{\sigma_s}{2 D_s}\left[2 + \left(\frac{i\omega}{-i(\omega + \tilde{B}) +D_sk^2}
+ (\tilde{B} \rightarrow -\tilde{B} )\right)\right]
\end{equation}
This then determines $\chi_{xx}^{''}(k, \omega)$ through the 
equality
\begin{equation}
\chi_{xx}^{''}(k, \omega) = 2\hbar Im \chi_{xx}(k, \omega).
\end{equation} 
The result is
\begin{equation}
\label{chiB}
\chi_{xx}^{''}(q, \omega) = \hbar\sigma_s \omega k^2\left[ \frac{1}{(\omega + \tilde{B})^2 + (D_sk^2)^2} + (\tilde{B}
 \rightarrow -\tilde{B})\right]
\end{equation}
Exactly the same expression
holds for $\chi_{yy}^{''}$ . 
Notice that the form of the spectral weight given in Eq.\ref{specwt} may 
be obtained from the above by setting the magnetic field $B$ to zero.

It is now possible to evaluate the change in free energy due to the magnetic field.
For details, see  Appendix \ref{AAC}.
The susceptibility is then obtained by differentiating with respect to 
the field. We find, for the interaction correction,
\begin{equation}
\Delta \chi_0 (T)  = \frac{b_{si}}{D_s}\left(\frac{k_BT}{2\pi\hbar D_s}\right)^{1/2}
\end{equation}
in three spatial dimensions.
The coefficient $b_{si}$ is given by
\begin{equation}
b_{si} = -\frac{\hbar \tilde{u_t}\chi_0 {\cal I}'}{8 \pi^{3/2}},
\end{equation}
where ${\cal I}' = \int_0^{\infty} \frac{dy}{\sqrt{y}}\frac{d}{dy}\left(\frac{y}{e^y -1}\right)$.

The full low temperature behaviour of the spin susceptibility is then given by
\begin{equation}
\chi(T) - \chi_0 = \frac{b_s}{D_s}\left(\frac{k_BT}{2\pi\hbar D_s}\right)^{1/2},  
\end{equation}
where again $b_s$ has contributions from both quantum interference
and interaction corrections:
\begin{equation}
b_s = b_{sq} + b_{si} .
\end{equation}
Note again that $b_{sq}$ is a universal constant while $b_{si}$ depends on
the interaction strength. (With anisotropic diffusion, $D_s$ is
replaced by $ \left(D_{\perp}^2D_{\parallel}\right)^{\frac{1}{3}}$).
Thus measurements of the temperature dependence
of the spin susceptibility can also be used to infer the relative importance of
interaction and quantum interference effects.
 
The simple calculations above demonstrate the existence of Altshuler-Aronov 
effects in the thermodynamic properties of the superconducting thermal metal\cite{note}. 
It is also possible to show that interaction effects lead to 
a suppression of the electron tunneling density of states. For the tunneling of a $d$ particle,
the result can be established in quite a  straighforward way along the lines of Ref. \cite{AALR}.
However, the physical tunneling process involves tunneling of an electron. In Appendix \ref{AAC}, 
we show that under certain further approximations, the tunneling density of states of 
electrons is the same as that of the $d$ particles. Thus, interaction effects lead to 
a suppression of the physical tunneling density of states in the superconducting thermal metal 
as well.

\subsection{Superconducting thermal insulator}
We now move on to consider the effects of interactions in the 
superconducting thermal insulator. In a phase with a gap in the quasiparticle
spectrum, weak interactions are irrelevant and lead to no significant
effects. (This is analogous to the insignificance of weak interactions in an
ordinary band insulator). So we consider the more interesting case
of a gapless superconducting thermal insulator . We argued in Section \ref{SSI} that,
in the absence of interactions, the disorder averaged density of states
 vanishes, generically as $|E|^2$, on approaching the Fermi energy.
Despite this, we show here that arbitarily weak
repulsive interactions lead to the formation of free paramagnetic moments. 
This result is quite analogous to what happens in a conventional Anderson 
insulator; arbitrarily weak repulsive short-ranged 
interactions lead to the formation of free paramagnetic
moments. 
(Note however that the density of states is a constant in the conventional Anderson insulator)

The discussion is
simplest in terms of the lattice Hubbard-type Hamiltonian Eqn. \ref{Hubb}.
As in Section \ref{SSI}, consider the case of strong on-site randomness and weak hopping. 
In the limit of zero hopping,  the resulting single site problem can be solved 
exactly even in the presence of the interaction term. To that end,
consider the single site Hamiltonian,
\begin{eqnarray}
{\cal H} & = & {\cal H}_0 + {\cal H}_U ,\\
{\cal H}_0 & = & t\sum_{\alpha}  c^{\dagger}_{\alpha}c_{\alpha} + 
\Delta c^{\dagger}_{\uparrow}c^{\dagger}_{\downarrow} +
\Delta^* c_{\downarrow}c_{\uparrow} ,\\
{\cal H}_U & = & U n(n-1),
\end{eqnarray}
with $n = \sum_{\alpha}  c^{\dagger}_{\alpha}c_{\alpha}$.
For the time being, we assume $U > 0$, making the interactions repulsive.  It is convenient 
to go to the $d$ representation:
\begin{eqnarray}
{\cal H}_0 & = & d^{\dagger}\left(t \sigma_z + \Delta_r \sigma_x + \Delta_i \sigma_y \right)d ,\\
{\cal H}_U & = & U \left[ (d^{\dagger} \sigma_z d)^2 + d^{\dagger}\sigma_z d \right].
\end{eqnarray}
Here $\Delta_{r,i}$ are the real and imaginary parts of $\Delta$. 
We will assume that  $t, \Delta_r, \Delta_i$ are random variables
with a probability distribution that has finite non-zero weight 
when all three variables are zero. As we argued in Section \ref{SSM},
in that case, the quasiparticle density of states of the non-interacting
Hamiltonian goes to zero as $|E|^2$. 

To diagonalize the full interacting Hamilonian,
note that the $d$ particle number is conserved even with interactions. Thus we may look for 
eigenstates with fixed $d$ particle number $n_d = \sum_{\alpha}d^{\dagger}_{\alpha}d_{\alpha}$. 
The physical spin
is determined entirely by $n_d$ (Eqn. \ref{spin}) through the relation
$S_z = \frac{1}{2}(n_d - 1)$. For the
single site problem, the Hilbert space consists of four states: 
$|0 \rangle$, $|\uparrow \rangle = d^{\dagger}_{\uparrow}|0 \rangle$,
$|\downarrow \rangle = d^{\dagger}_{\downarrow}|0 \rangle$,
and $|\uparrow \downarrow \rangle =  d^{\dagger}_{\uparrow}d^{\dagger}_{\downarrow}|0 \rangle$.
The states $|0\rangle$ and $|\uparrow \downarrow \rangle$ are immediately 
seen to be eigenstates with energy $0$. Diagonalizing the Hamiltonian 
in the space of the remaining two states $|\uparrow \rangle$, $|\downarrow \rangle$,
we find  two other eigenstates $|+\rangle, |-\rangle$ with the eigenvalues
$E_{\pm} = U \pm \sqrt{(t+U)^2 + |\Delta|^2}$. Note that the lower of the two eigenvalues
$E_{-}$ is negative if $-2tU < t^2 + |\Delta|^2$. (The other eigenvalue $E_{+}$ is 
always positive). If $E_{-} < 0$, then the ground state is the state $|-\rangle$. 
This lies in the subspace with $n_d = 1$, and hence has physical spin $= 0$. 
If $E_{-} > 0$, then the ground state is two fold degenerate with both $|0\rangle$,
and $|\uparrow \downarrow \rangle$ having zero energy. These correspond to states where the 
ground state has a physical spin $1/2$. 

Thus a free magnetic moment is formed in the ground state of a single site 
whenever the condition $t^2 + |\Delta|^2 < -2tU$ is satisfied. For a collection  
of sites with a generic random distribution  $P(t, \Delta_r \Delta_i)$
of $t, \Delta_r$, and  $\Delta_i$, there will always be some weight where this condition
is satisfied. Hence there will be a finite probability of forming a 
free moment at any site. This leads to a finite density of magnetic
moments for the full system.

It is natural to expect that inclusion of weak hopping between sites
does not change the result above, so long as we are in the localized phase. 
So we conclude that arbitrary weak repulsive interactions lead to the formation of 
free paramagnetic moments in the superconducting thermal insulator .

What is the ultimate fate of these magnetic moments at low energies? There are two generic
possibilities: the spins can freeze into a spin glass phase, or stay unfrozen in a 
phase with random singlet bonds\cite{RS} between pairs of spins. It is 
a difficult matter to decide on the conditions for realizing either
of these possibilities, and we will not attempt it here.

The analysis above has assumed that the interaction was repulsive. If the interaction is attractive, {\em i.e} $U < 0$,
the solution of the single site problem shows that the ground state has no net spin. Thus there is no
local moment instability in that case. The spin susceptibility vanishes at zero temperature, as
in the non-interacting case.

\section{Systems with time reversal invariance}
\begin{figure}
\epsfxsize=3.5in
\centerline{\epsffile{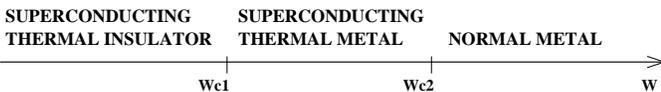}}
\vspace{0.15in}
\caption{Zero temperature phase diagram for the superconductor-normal metal
transition in the presence of time reversal invariance. 
The parameter $W$ is a measure of the strength of the disorder.}
\vspace{0.15in}
\label{SNMT}
\end{figure}

\begin{figure}
\epsfxsize=3.5in
\centerline{\epsffile{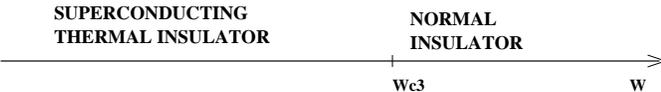}}
\vspace{0.15in}
\caption{Zero temperature phase diagram for the superconductor-insulator
transition in the presence of time reversal invariance.}
\vspace{0.15in}
\label{SIT}
\end{figure}

In this section, we will briefly consider systems with time reversal invariance.
An $s$-wave superconductor with weak or moderate impurity scattering has a gap
in the quasiparticle excitation spectrum, and hence is in the 
superconducting thermal insulator phase.  Strong 
impurity scattering will destroy the superconductor. When this happens, the resulting phase is
most likely an insulator. However we see no reason of principle 
forbidding a transition to a normal metal (in three spatial dimensions) 
when the superconductor is destroyed. We propose that the transition then occurs from 
the superconducting thermal metal phase, {\em i.e} as the impurity strength is increased,
there is a transition from the superconducting thermal insulator to the superconducting thermal metal
which precedes the ultimate transition to the normal metal (See
Figure \ref{SNMT}). On the other hand, if the transition
is to the insulator, we propose that it is directly from the superconducting thermal 
insulator (Figure \ref{SIT}).

The properties of the superconducting thermal metal phase in this case (with time reversal symmetry)
are quite similar to the case where time reversal symmetry is absent. Differences exist, however,
in the superconducting thermal insulator phase. The heuristic argument\cite{dos}, in the beginning of 
Section \ref{SSI}, for the quasiparticle density of states now shows that it 
vanishes in this case as well,
but only as fast as or faster than $|E|$. As in Section \ref{SSI}, we will 
provide supporting numerical
evidence for this statement by calculating the density of states of the Bogoliuobov-deGennnes
equations appropriate for a time reversal invariant system in one dimension. 
In order to preserve time reversal symmetry, we set the imaginary part 
of the gap function $\Delta^{(2)}$ in the Hamiltonian of Eqn.\ref{ham} of 
Section \ref{SSI} to zero. Apart from this important difference, 
the models that we use here are the same as those of Section \ref{SSI}. 
Again, we first display results for a one dimensional model where 
there is a gap in the clean limit.

\begin{figure}
\epsfxsize=3.5in
\centerline{\epsffile{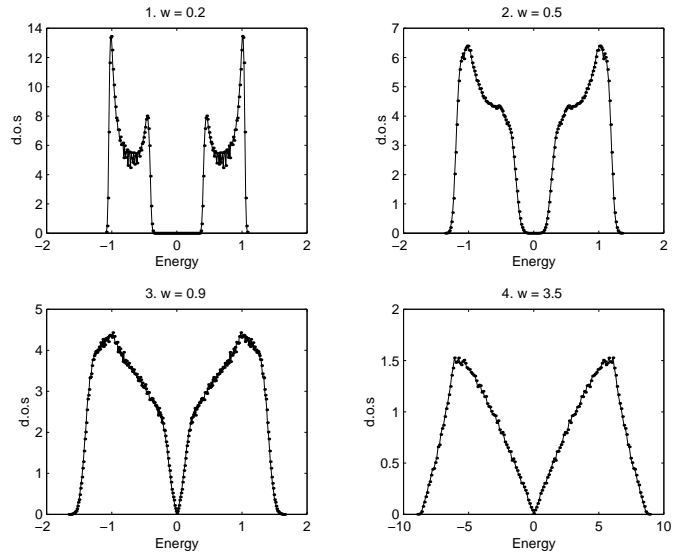}}
\vspace{0.15in}
\caption{Density of states in the $T$ -invariant 
superconducting thermal insulator  showing evolution with increased disorder, W. The energy 
scale is set by the bandwidth of the pure dispersion of Eqn. \ref{purgap} of Section \ref{SSI}.}
\vspace{0.15in}
\label{Tgap}
\end{figure}

\begin{figure}
\epsfxsize=3.5in
\centerline{\epsffile{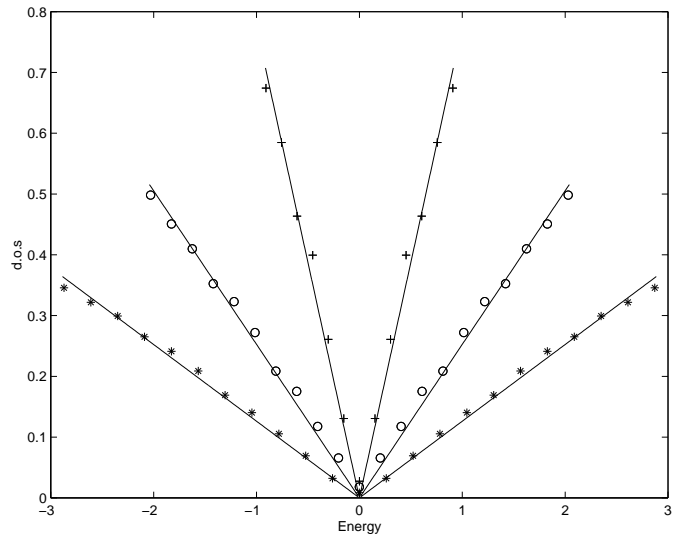}}
\vspace{0.15in}
\caption{Density of states in the time reversal -invariant 
superconducting thermal insulator  at low energies. Decreasing slope corresponds to increasing disorder strength, W, that ranges from 0.8 to 4.}
\vspace{0.15in}
\label{zoomT}
\end{figure}

The density of states clearly vanishes at zero energy. Closer examination
of the low energy behaviour(See Figure \ref{zoomT}) shows that it actually vanishes
as $|E|$ in agreement with the simple argument above. 

\begin{figure}
\epsfxsize=3.5in
\centerline{\epsffile{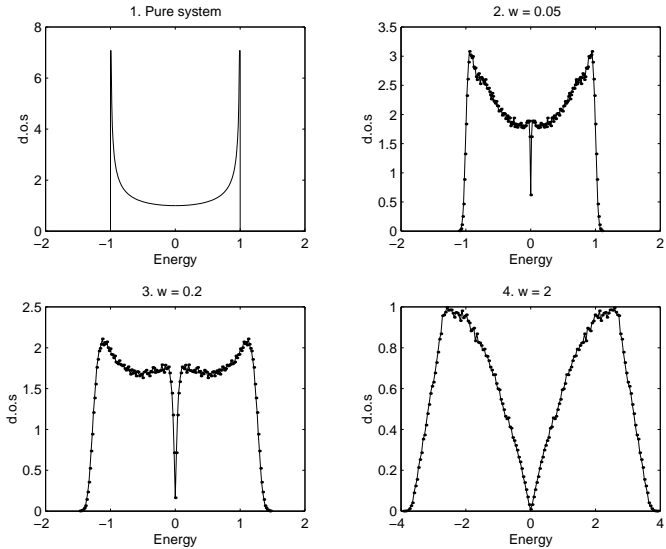}}
\vspace{0.15in}
\caption{Evolution of density of states with disorder, W, in model where it 
is constant in the clean limit. Here, the energy scale is set by the 
bandwidth of the pure dispersion given by Eqn.\ref{purconst} of Section \ref{SSI}. }
\vspace{0.15in}
\label{Tconst}
\end{figure}

We also considered the situation 
where the density of states is a constant in the clean limit. The results are 
shown in Figure \ref{Tconst}.
Again, the density of states vanishes linearly for strong disorder.

\section{Phase transitions}
\begin{figure}
\epsfxsize=3.5in
\centerline{\epsffile{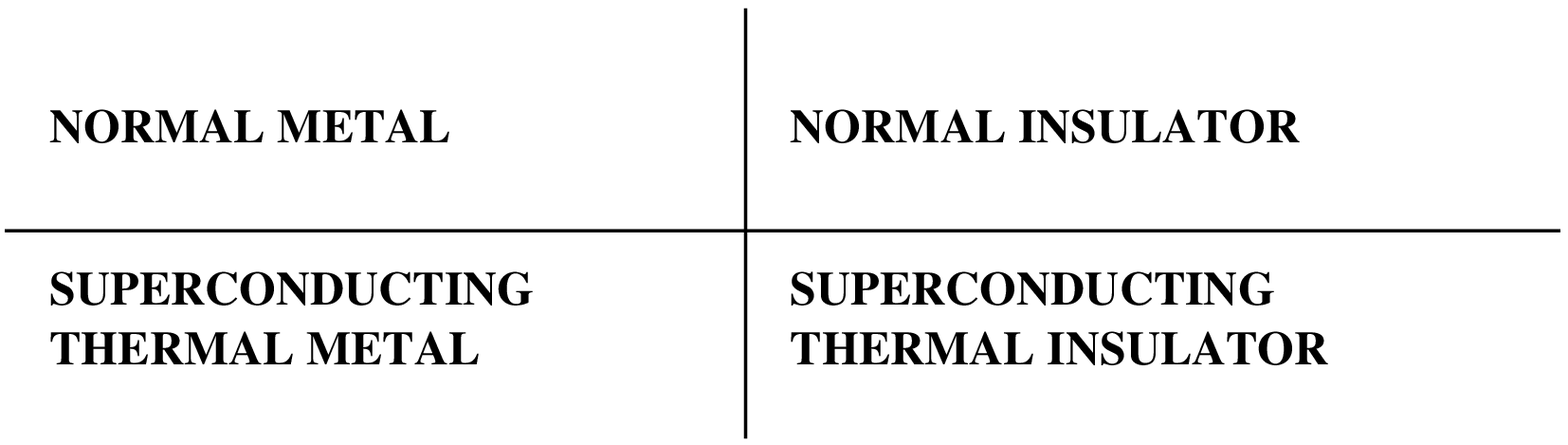}}
\vspace{0.15in}
\caption{Zero temperature phase diagram showing transitions where the 
superconducting phase is destroyed}
\vspace{0.15in}
\label{multcrit}
\end{figure}

In this section, we study some general properties of the various transitions
between the phases that we have discussed (see Fig.\ref{multcrit}. The most novel phase transition
in our phase diagram is the one between the superconducting thermal metal
and the superconducting thermal insulator. This is a localization transition
that occurs {\em inside} the superconducting phase, and should be accessible
experimentally. On approaching the transition from the thermal metal side,
the spin conductivity goes to zero continuously. Similarly, the 
low temperature ratio $A=\kappa/T$, which is non-zero in the thermal metal, vanishes at the transition.
The low temperature thermal transport in the superconducting thermal insulator is
presumably through variable-range hopping, and 
$\kappa \sim e^{- \left(\frac{T_0}{T}\right)^{x}}$ with $x = 1/2$.
In the critical region near the transition point, various physical quantities 
are expected to have universal singular behaviour.
In the model of non-interacting quasiparticles, 
it is  possible to develop systematic 
calculations of the universal critical exponents\cite{S5}. 
However, we expect universal properties even with interactions present. 

On approaching the transition at zero temperature by varying the 
field strength towards $H_{c4}$, there is a length scale $\xi$
(which may be interpreted as the quasiparticle localization length in the localized
phase) that diverges as
\begin{equation}
\label{xih}
\xi \sim |H - H_{c4}|^{-\nu}.
\end{equation}
Similarly, moving away from the critical point by turning on a finite temperature 
introduces a length scale $\xi_T$ which behaves as
\begin{equation}
\label{xit}
\xi_T \sim T^{-\frac{1}{z}}
\end{equation}
Note that the two equations (\ref{xih}) and (\ref{xit}) define the two critical
exponents $\nu$ and $z$.

On approaching the transition point from the delocalized side, the coefficient 
$A$ introduced above goes to zero. 
Precisely at the transition point $H = H_{c4}$, but at finite temperature,  
$\kappa(T) \sim T^{1+\phi}$
with $\phi > 0$ being a universal exponent. 
In general, for fields close to $H_{c4}$ and low temperatures, we may write down
a scaling form
\begin{equation}
\label{scale}
\frac{\kappa }{T}  =  c_1T^{\phi} \tilde{Y}\left(\frac{\xi_T}{\xi} \right) .
\end{equation}
The constant $c_1$
is non-universal, while the function $\tilde{Y}$ is universal. 
Equivalently, we may use Eqns. (\ref{xih}) and (\ref{xit})
to write 
\begin{equation}
\frac{\kappa }{T}  = c_1 T^{\phi}Y\left(c_2\frac{|H- H_{c4}|}{T^{\frac{1}{z\nu}}}\right) .
\end{equation}
Here $c_2$ is also non-universal, and $Y$ is  
a universal scaling function such that $Y(0)$ is a finite constant. Further,
requiring that $\kappa \sim T$ as $T \rightarrow 0$ on the metallic side, we see that
$Y(x \rightarrow +\infty) \sim x^{z\nu\phi}$. This 
then implies that the coefficient $A$ introduced above vanishes, on 
approaching $H_{c4}$, as $(H - H_{c4})^{z\nu\phi}$. 
Further, if we make the assumption that the fixed point theory controlling the 
transition obeys hyperscaling, then conventional scaling arguments can be
used to show the exponent equality
\begin{equation}
\phi = \frac{d -2}{z}.
\end{equation}
Similar scaling forms, but with different exponents and scaling functions,
describe the transition in the time reversal invariant case as well.

Our phase diagram has important implications for the phase transitions 
at zero temperature where 
the superconducting phase is destroyed. For the superconductor-normal metal transition,
we suggest that the transition is generically from the superconducting thermal 
metal phase, {\em i.e}, from a superconductor with spin and energy diffusion at $T = 0$.
Further this implies that the superconductor-normal transition generically occurs from a 
gapless superconductor with a finite, non-zero density of quasiparticle states at the Fermi
energy. The presence of gapless quasiparticle excitations in the superconducting side 
should affect strongly the critical properties of 
the SC-normal metal transition: previous theoretical 
treatments of this transition\cite{KB} which have ignored this 
feature are hence expected to be incorrect. 
Similarly, the superconductor-insulator transition is generically 
from the superconducting 
thermal insulator.

As shown in Fig. \ref{multcrit}, direct transitions from the superconducting thermal metal to the 
normal insulator or from the superconducting thermal insulator to the normal metal are possible
at a special multicritical point. We will however not attempt to describe the properties 
of such a multicritical point here.

\section{Discussion}
One of the main purposes of this paper is to point out that {\em all} superconductors fall 
into one of two categories - those that, apart from being superconducting, share many
properties with conventional metals, and those that share many properties with conventional
insulators. This distinction can be made quite precise, and indeed corresponds to 
the existence of two distinct zero temperature superconducting phases which are 
distinguished by the nature of quasiparticle transport.  In this final section, we will discuss various  
experimental  implications. This will be followed by a discussion of various peripheral 
matters that have been omitted from our considerations so far, and their effects on 
experimental systems.

\subsection{Experiments}
 A powerful way of 
probing quasiparticle transport in a superconductor is through
thermal conductivity measurements.   
In the superconducting ``insulator" phase, the ratio $\kappa/T$ goes to 
zero as the temperature goes to zero. On the other hand, in the superconducting ``metal",
$\kappa/T $ goes to a constant as the temperature goes to zero. 

The Type $II$
$s$ -wave superconductor offers a definite opportunity for tuning between these two
phases. At low fields, this is in the superconducting ``insulator" phase.
Consequently, $\kappa/T \rightarrow 0$ as $ T \rightarrow 0$. Upon increasing the field, 
there could, under conditions we have outlined,  
be a delocalization transition to the superconducting ``metal" phase with 
$\kappa/T$ going to a constant. It should also be possible to explore the properties 
of the phase transition between the superconducting ``insulator" and the superconducting 
``metal". Right at the critical point, the thermal conductivity $\kappa \sim T^{1+\phi}$ with
$\phi > 0$. We are not aware of any experimental investigations of this novel ``metal-insulator"
transition inside the superconducting phase so far. In performing  the heat transport measurements,
it is, of course, necessary to ensure that the 
phonon contributions have been subtracted out\cite{note1}. 

Another interesting experimental possibility is provided by three dimensional dirty $s$-wave superconductors 
in the absence of any external magnetic fields. Upon varying the impurity concentration	to destroy
superconductivity, if the transition is to a normal metal, we have proposed that it occurs from the 
superconducting ``metal" phase. As the superconductor is in the superconducting ``insulator"
phase at low impurity concentrations, there will then be a phase transition to the superconducting ``metal" 
with increasing impurity concentration before the destruction of superconductivity. This too
can be probed by thermal transport measurements. The phase transition in this case is similar to
the ``metal-insulator" transition discussed above, but belongs to a different universality 
class due to the presence of time reversal symmetry.

Recent measurements\cite{Taill} of the low temperature in-plane 
thermal conductivity in the high temperature superconductors
show that $\kappa/T$ goes to a constant as $T \rightarrow 0$. Whether this is indicative of 
a true three dimensional superconducting ``metal" phase stabilized by
interlayer quasiparticle hopping, or just a two dimensional weakly 
localized superconducting ``insulator" (see below) is difficult to ascertain at present.

A number of experimental predictions follow from our study of the properties of
either phase. We start with the superconducting ``metal". For systems with 
negligible spin-orbit scattering, this phase is characterized by the presence 
of spin diffusion at zero temperature. Measurements of the spin diffusion
constant should then show a non-zero value at zero temperature. The spin susceptibility,
as measured, for instance, by the Knight shift is also predicted to 
saturate to a finite, non-zero value at zero temperature. Note however that
due to quantum interference, and interaction effects, we predict a $\sqrt{T}$
dependence in the temperature dependence of the spin susceptibility at low temperatures.
Similarly the specific heat has a $T^{3/2}$ correction. As we emphasized in Section \ref{Int},
either the specific heat or the susceptibility measurements may be used to
quantify the relative importance of interaction and quantum interference effects.

In the superconducting thermal insulator  phase, the spin conductivity is zero at zero temperature.
Further, if the interactions are weak, then the vanishing density of states in the 
non-interacting quasiparticle model manifests itself in all thermodynamic properties.
For instance, the specific heat vanishes faster than $T^{\alpha+1}$ with $\alpha = 1,2$ depending on
whether time reversal is a good symmetry or not. Similarly, the spin susceptibility vanishes
faster than $T^{\alpha}$. Interaction effects become important at the lowest temperatures
- formation of free paramagnetic moments would initially give rise to a Curie term in the susceptibility 
which would then be altered at even lower temperatures due to exchange between these moments.

So far, we have neglected the Zeeman coupling and spin orbit interactions. In making contact with experiments, 
it is essential to have some understanding of the effects  
of including these. We discuss that next.

\subsection{Zeeman coupling}
Weak Zeeman coupling does not affect the existence of the two kinds of
superconducting phases, but modifies their properties. For a magnetic field along, the 
$z$-direction of spin, the components of spin along the $x,y$ axes are no longer conserved.
This cuts off their diffusion at long length and time scales in the superconducting ``metal''  phase. 
The $z$-component of the 
spin and the energy continue to diffuse though. Further, both the quantum interference
and Altshuler-Aronov interaction corrections to the thermodynamic quantities in the ``metallic"
phase are cut-off by the finite Zeeman coupling. 
For instance, this leads to a field
dependent spin susceptibility at the lowest temperatures of the form
\begin{equation}
\chi(B) - \chi_0 \sim B^{1/2}.
\end{equation}
For the interaction correction, this is demonstrated in the Appendix. For the quantum interference
contribution, this result may be understood by noting that the Zeeman magnetic field acts as 
a ``chemical potential" for the $d$-particles (as it couples to the physical spin $\sim$
$d$-particle number). The density of states at the Fermi energy (and hence the spin susceptibility)
is then given by Eqn. \ref{rhossm} to be $\sim B^{1/2}$. 
On the spin-insulating side, in the approximation of ignoring interactions, the effect of 
finite Zeeman 
coupling depends on whether or not a hard gap exists in the quasiparticle excitation
spectrum. If gapped, weak Zeeman coupling is innocuous, and can be ignored. On the other hand,
if the system is gapless with a density of states vanishing as $|E|^{\alpha}$ ($\alpha = 1,2$
depending on whether time-reversal is a good symmetry or not), then weak Zeeman coupling
leads to a finite density of states at the Fermi energy, proportional
to $|B|^{\alpha}$.

\subsection{Spin-orbit effects}
Spin-orbit scattering, like Zeeman coupling, does not affect the existence of the two kinds of superconducting
phases in three dimensions, but modifies their properties. In the presence of spin-orbit scattering,
no component of the spin diffuses in the superconductor with delocalized quasiparticles. 
Energy continues to diffuse in this phase however. 
Thus this phase is characterized by a finite value of $\kappa /T$ as $T \rightarrow 0$. In contrast
to systems with conserved spin, the density of states in the non-interacting quasiparticle theory 
has a $\sqrt{E}$ {\em enhancement}\cite{SO} at low energies. Similarly, in the localized insulator, 
in the non-interacting theory, spin-orbit effects allow\cite{SO} for the possibility of a finite density of 
quasiparticle states at the Fermi energy.

\subsection{Two dimensional systems}
Through out this paper, we have focused on three dimensional systems. Here we 
make some brief remarks about two dimensional systems. For systems with conserved spin,
quantum interference effects lead\cite{short} to an absence of spin and energy diffusion at the 
longest length scales. The superconducting thermal metal therefore does not exist in two dimensions,
at least if quasiparticle interactions are ignored. The superconductor-insulator transition in
two dimensions therefore occurs from the superconducting thermal insulator phase. 
Spin-orbit scattering can stablilize\cite{Zirn,SO} a phase    
with delocalized quasiparticles in the two dimensional superconductor in the approximation of
ignoring quasiparticle interactions. The resulting phase has a divergent $\kappa /T$ as $T \rightarrow 0$,
and a divergent density of quasiparticle states at the Fermi energy\cite{SO}.
 
This research was supported by NSF Grants DMR-97-04005,
DMR95-28578
and PHY94-07194.

\appendix
\section{Interaction Hamiltonian}
\label{intham}
Here, we provide some microscopic justification for the interaction Hamiltonian Eqn. \ref{Hint}. 
For simplicity, we consider a clean system in $d =2$. This may be described by the model Hamiltonian
\begin{eqnarray}
{\cal H} & = & {\cal H}_0 + {\cal H}_{coul} + {\cal H}_{el-ph} ,\\
{\cal H}_0 & = & \int d^2x \sum_{\sigma}\psi^{\dagger}_{\sigma}\left(-\frac{\vec{\nabla}^2}{2m} - \mu \right)\psi_{\sigma} ,\\
{\cal H}_{coul} & = & \sum_{\sigma, \sigma'}\int_{x,x'} \psi^{\dagger}_{\sigma}(x)\psi_{\sigma}(x)V(x-x')
 \psi^{\dagger}_{\sigma'}(x')\psi_{\sigma'}(x') .
\end{eqnarray}
Here $V(x-x') \sim \frac{1}{|x-x'|}$ is the Coulomb interaction. ${\cal H}_{el-ph}$ is the electron-phonon
interaction which we do not specify in detail other than to assume that it leads to an effective attractive interaction
at energies smaller than the Debye frequency $\omega_D << E_F$  about the Fermi energy. 
We first imagine integrating out all modes\cite{Shankar} except those within  $\omega_D$ 
about the Fermi energy. In the resulting low energy theory, only a small number of qualitatively different
interaction amplitudes are allowed due to geometrical restrictions imposed 
on the scattering processes\cite{Shankar}. These correspond to a  
charge density-charge density (``singlet" channel), spin density-spin density
(``triplet" channel) interactions and to two interactions between spin singlet and spin triplet 
Cooper pair operators. Perturbative renormalization group arguments\cite{Shankar} show that the 
singlet and triplet amplitudes are marginal up to one loop,
while the Cooper channel amplitudes are marginally relevant if attractive. We assume such an 
attractive interaction leads to a flow towards a spin-singlet BCS superconductor. Treating the 
attractive spin-singlet Cooper interaction in mean field theory, and ignoring the other interactions 
in the singlet and triplet channels is equivalent to conventional BCS mean field theory. Going beyond
the BCS theory requires reinstating the singlet and triplet interactions, and including fluctuations
of the BCS order parameter. The order parameter fluctuations may be integrated out in the 
superconducting phase leading to an effective four fermion interaction which renormalizes the singlet 
amplitude. This then leads to an effective action for the quasiparticles in the superconductor
which includes quasiparticle interactions.

Consider now the system we have focused on the most - the Type $II$ superconductor in a field
in the presence of disorder. In principle, the interaction Hamiltonian contains both the singlet
(charge-density) interactions and the triplet (spin-density) interactions. However, the charge density
is not a hydrodynamic mode in the superconductor, and does not diffuse. Consequently, the singlet interaction
is expected to be quite innocuous. We therefore retain only the spin triplet interaction. 

The sign of the (bare) triplet interaction $u_t$ is determined 
by the balance between the repulsive Coulomb interaction amplitude in the triplet
channel and the attractive (phonon) interaction amplitude in the same channel. This is,
in principle, different from the balance in the Cooper channel where a net attractive interaction is
required for superconductivity. Thus, $u_t$ can be either positive or negative. It can be shown that
$u_t < 0$ corresponds to repulsive interactions.

\section{Altshuler-Aronov corrections}
\label{AAC}
In this appendix, we provide details of the calculation of the singular corrections
to the properties of the superconducting thermal metal  phase due to interactions.

\subsection{Specific heat correction}
As shown in Section \ref{Int}, the correction to the free energy at zero magnetic field
is given by
\begin{displaymath}
\Delta {\cal F} = 6 \hbar \sigma_s \int d\vec k\frac{d\omega}{2\pi}
u_t(k)\left(\frac{1}{1- e^{-\beta\hbar \omega}}\right)\frac{\omega k^2}{\omega^2 + D_s^2 k^4} ,
\end{displaymath}
where $u_t(k) = \int d^3x ~e^{-ik \cdot x}u_t(x)$ is the Fourier transform of the 
triplet interaction $u_t(x)$. The $k$ integral is restricted to $|k| < \Lambda \sim \frac{1}{l_e}$
where $l_e$ is the elastic mean free path. (We have denoted $d\vec k \equiv \frac{d^3 k}{(2\pi)^3}$).
The $\omega$ integral runs from $-\infty$ to $+\infty$.
If the range of the short-ranged interaction $u_t(x)$ is much smaller than the mean free path (which we
assume), then the Fourier transform $u_t(k)$ may be approximated by it's value at $k =0$,
{\em i.e} $u_t(k) \approx \tilde{u}_t = \int d^3x u_t(x)$. Subtracting out the zero temperature
correction to the free energy, we get
 
\begin{eqnarray*}
\delta_T \Delta{\cal F} & \equiv & \Delta {\cal F}(T) - \Delta {\cal F}(0), \\
\frac{\delta_T \Delta{\cal F}}{6\hbar \tilde{u}_t \sigma_s} & = &  \int d\vec k\frac{d\omega}{2\pi}
\left(\frac{1}{1- e^{-\beta\hbar \omega}} - \theta(\omega)\right)\frac{\omega k^2}{\omega^2 + D_s^2 k^4} \\
& = & \frac{1}{2\pi^3} \int_0^{\Lambda} dk k^4 \int_0^{\infty}d \omega
\left(\frac{1}{e^{\beta\hbar \omega} - 1}\right) \frac{\omega}{\omega^2 + D_s^2 k^4} .
\end{eqnarray*}
 
In going to the second line, we have integrated over the angular coordinates of the momentum $\vec k$,
and have reduced the frequency integral to one over positive $\omega$ alone. We now make the change of variables
$y = \beta \hbar \omega$, $x = \sqrt{\beta \hbar D_s} k$ to get
\begin{displaymath}
\frac{\pi^3 \delta_T (\Delta{\cal F})}{3\hbar \tilde{u}_t \sigma_s} =  
\left(\frac{k_B T}{\hbar D_s}\right)^{\frac{5}{2}}\int_{x,y} x^4  \left(\frac{y}{e^y -1}\right)
\left(\frac{1}{y^2 + x^4}\right) .
\end{displaymath}
Here $\int_{x,y} \equiv \int_0^{x_0} dx \int_0^{\infty} dy$ and 
$x_0 = \Lambda \sqrt{\frac{\hbar D_s}{k_B T}}$. The integral over $x,y$ above can clearly
be rewritten as
\begin{displaymath}
\frac{\pi^2}{6} x_0 - \int_0^{x_0} dx \int_0^{\infty} dy  \left(\frac{y^3}{e^y -1}\right)
\left(\frac{1}{y^2 + x^4}\right).
\end{displaymath}
The first term contributes ${\cal O}(T^2)$ to the free energy, and is hence non-singular. 
All the singular corrections come from the second term. As the $x$-integral is ultraviolet convergent
in this term, we set $x_0 = \infty$ and evaluate it explicitly to obtain the singular correction 
to the free energy:
\begin{equation}
\Delta {\cal F}(T) - \Delta {\cal F}(0) = -\frac{9\hbar\tilde{u}_t \sigma_s}{4 (2\pi)^{3/2}}\zeta(5/2)
\left(\frac{k_B T}{\hbar D_s}\right)^{5/2} .
\end{equation}
Expressing $\sigma_s$ in terms of $D_s$ and $\chi_0$ using the Einstein relation Eqn. \ref{ES}, and 
differentiating with respect to $T$ to get the specific heat, we get the result quoted in Section \ref{Int}.

\subsection{Susceptibility correction}
The field dependent term in the correction to the free energy is 
\begin{displaymath}
\Delta{\cal F}(B) = 2\hbar \tilde{u}_t \int d\vec k  \frac{d\omega}{2\pi}
\frac{1}{1-e^{-\beta \hbar \omega}}\chi_{xx}^{''}(k, \omega) .
\end{displaymath}
The factor of two in front accounts for the contribution from both the $S_xS_x$ and $S_yS_y$
correlators, and $\chi_{xx}^{''}(k, \omega)$ is given by Eqn. \ref{chiB}. To calculate the susceptibility,
we imagine evaluating the free energy in a large, finite box of linear size $L$. We differentiate with respect 
to the field, and take $B \rightarrow 0$ to get the susceptibility. The limit $L \rightarrow \infty$ is 
taken at the end. This gives
\begin{eqnarray*}
\frac{\Delta \chi(T)}{2\hbar \sigma_s\tilde{u}_t (g\mu_B)^2} & = & -\int d\vec k\frac{d\omega}{2\pi}
\frac{\omega k^2}{1-e^{-\beta \hbar \omega}}\frac{\partial^2}{\partial \omega^2}\frac{1}{\omega^2 + D_s^2 k^4} \\
& = & -\int d\vec k\frac{d\omega}{2\pi}\frac{\partial^2}{\partial \omega^2}\left(\frac{\omega k^2}{e^{\beta \hbar \omega}-1}\right)
\frac{1}{\omega^2 + D_s^2 k^4}
\end{eqnarray*}
It is assumed that the $k$-integral is cut-off at the lower end by the inverse system size $L^{-1}$ 
and at the upper end by the inverse mean free path. In going to the second line, we have performed an integration
by parts twice. We may now proceed exactly as for the specific heat correction above. We first replace the 
frequency integral by one that runs over positive $\omega$ alone, and integrate over the angular components of the 
momentum to get
\begin{displaymath}
\frac{\pi^3 \Delta \chi(T)}{\hbar\tilde{u}_t (g\mu_B)^2 \sigma_s} = - \int_{k, \omega} k^4
\frac{\partial^2}{\partial \omega^2}\left(\frac{\omega}{e^{\beta \hbar \omega}-1}\right)
\frac{1}{\omega^2 + D_s^2 k^4}
\end{displaymath}
(We denote $\int_{k,\omega} = \int_{L^{-1}}^{\Lambda} dk \int_0^{\infty}d \omega$). 
The $k$-integral is infra-red convergent, and may be performed first as before. The singular 
contribution may be evaluated exactly as for the specific heat above, and yields the result 
quoted in Section \ref{Int}. 

\subsection{Electron tunneling density of states}
Here we show that the electron tunneling density of states is, under certain approximations,
essentially the same as the $d$-particle tunneling density of states. For concreteness,
we consider a lattice electron Hamiltonian. The 
single-particle electron density of states($N_c(E)$) is related to the electron Green's function 
${\cal G}$ by
\begin{equation}
N_c(E) = -\frac{1}{\pi} Im\overline{{\cal G}_{ii}(E + i\eta)} ,
\end{equation}
where the overline denotes averaging over the disorder and $i$ is a site index on the lattice.
The Green's function ${\cal G}$ may be obtained by analytic continuation from imaginary frequencies:
\begin{equation}
{\cal G}_{ii}(i\omega)  = \langle c_{i\uparrow}(\omega)\bar{c}_{i\uparrow}(\omega) + 
(\uparrow \rightarrow \downarrow) \rangle .
\end{equation}
Transfoming to the $d$-fields, the right-hand side becomes
\begin{displaymath}
\langle d_{i\uparrow}(\omega)\bar{d}_{i\uparrow}(\omega) - 
d_{i\downarrow}(-\omega)\bar{d}_{i\downarrow}(-i\omega) \rangle .
\end{displaymath}
Defining the $d$-particle Green's function $G_{ii, \alpha \alpha}(i\omega) = 
\langle d_{i\alpha}(\omega)\bar{d}_{i\alpha}(\omega) \rangle$, this becomes
\begin{displaymath}
G_{ii, \uparrow \uparrow}(i\omega) - G_{ii, \downarrow \downarrow}(- i\omega) .
\end{displaymath}
Now spin $SU(2)$ invariance requires $c_{i\uparrow}(\omega)\bar{c}_{i\uparrow}(\omega)
= c_{i\downarrow}(\omega)\bar{c}_{i\downarrow}(\omega)$ which implies that
\begin{equation}
\label{dgsu2}
G_{ii, \uparrow \uparrow}(i\omega) = - G_{ii, \downarrow \downarrow}(- i\omega) .
\end{equation}
We therefore have 
\begin{equation}
\label{dcsu2}
{\cal G}_{ii}(i\omega) = 2G_{ii,\uparrow\uparrow}(i\omega) .
\end{equation} 
The $d$-particle tunneling density of states is 
\begin{equation}
N_d(E) = -\frac{1}{\pi} Im\left(\overline{G_{ii, \uparrow\uparrow}(E+i\eta) + 
G_{ii,\downarrow\downarrow}(E+i\eta)}\right) .
\end{equation}
Using Eqns. \ref{dgsu2} and \ref{dcsu2}, we may rewrite this as
\begin{eqnarray}
N_d(E) & = & -\frac{1}{2\pi} Im \left(\overline{{\cal G}_{ii}(E +i\eta) - {\cal G}_{ii}(-E - i\eta)}\right) \\
& = & \frac{1}{2}\left(N_c(E) + N_c(-E) \right) .
\end{eqnarray}
If we now finally assume that asymptotically close to the Fermi energy, there is a statistical particle-hole
symmetry for the electrons, then $N_c(E) \approx N_c(-E)$ as $E \rightarrow 0$. We then have
$N_c(E) \rightarrow N_d(E)$ as $E \rightarrow 0$.

The methods of Ref. \cite{AALR} can be used to show in a straightforward way the existence of a $\sqrt{E}$ singularity 
due to interactions in $N_d(E)$ in the superconducting thermal metal  in three dimensions. As we noted in Section \ref{SSM},
$N_d(E)$ has singularities due to quantum interference as well. We therefore have, for the electron density
of states $N_c(E)$, at low energies
\begin{equation}
N_c(E) - N_c(0) \sim \sqrt{E} ,
\end{equation}
with contributions from both quantum interference and interaction effects.

\end{multicols}

\begin{references}
\bibitem{note0} This result assumes the conservation of spin; in the presence of spin
orbit coupling, it is possible\cite{Zirn,SO} 
to have a delocalized phase in two dimensions analogous to
corresponding results in normal metals.
\bibitem{note}Interestingly, arguments similar to the ones we have made 
provide an alternate route to understanding the diagrammatic calculations of Ref. \cite{AA}
for the interaction corrections to the thermodynamics of disordered 
normal metals (at least for the case of short-ranged interactions).
\bibitem{note1}Note however that at low temperatures, the phonon heat conductivity $\kappa _{ph}
\sim T^3$ goes to zero rapidly.
\end{references}
\end{document}